\documentclass[aps,prl,reprint,groupedaddress]{revtex4-1}
\usepackage{graphicx}  

\begin{document}


\title{
Nonconventional odd denominator fractional quantum Hall states \\
in the second Landau level}

\author{A. Kumar}
\author{M.J. Manfra}
\author{G.A. Cs\'{a}thy} 
\email[]{gcsathy@purdue.edu}
\affiliation{Purdue University, West Lafayette, Indiana 47907 } 

\author{L.N. Pfeiffer }
\author{K.W. West}
\affiliation{Princeton University, Princeton, New Jersey 08544} 


\date{\today}

\begin{abstract}
We report the observation of a new fractional quantum Hall state in the second Landau level of a two-dimensional electron gas at the Landau level filling factor $\nu=2+6/13$. 
We find that the model of noninteracting composite fermions can explain the magnitude of gaps of
the prominent $2+1/3$ and $2+2/3$ states. The same model fails, however, to account for the
gaps of the $2+2/5$ and the newly observed $2+6/13$ states suggesting that these two states are
of exotic origin.

\end{abstract}
\maketitle

Fractional quantum Hall states (FQHS) are incompressible electron liquids which form at rational 
ratios ($\nu$) of the number of electrons to the magnetic flux quanta penetrating a  
two-dimensional (2D) electron gas.
At large enough magnetic fields $B$ all electrons occupy the lowest Landau level (LL)
and major sequences of FQHS form at LL filling factors $\nu$ with odd denominators 
\cite{tsui,laughlin,jainCF,halp93}.  
Of these states the parent FQHS at $\nu=1/3$ and $1/5$ are described by Laughlin's wavefunction \cite{laughlin} 
while the numerous daughter states
such as the $\nu=2/5,3/7,4/9,5/11,...$ FQHS can all be understood 
within Jain's model of noninteracting composite fermions (MNCF) \cite{jainCF}. 
In the literature these states are often referred to as the composite fermion
(CF) hierarchy states, or the Jain states.
The MNCF maps the interacting electrons into noninteracting CF 
by the attachment of an even number of magnetic flux quanta to each 
electron and interprets the observed FQHS as being integer quantum Hall states of the composite particles. 
The MNCF not only accounts for the hierarchy of FQHS observed in the lowest LL but 
also specifies the relative strength of their energy gaps \cite{halp93,du93,mano94,pan00}.

The $\nu=2+1/2=5/2$ even denominator FQHS \cite{willett87} does not belong to CF hierarchy. 
Its existence indicated for the first time
that electron correlations in the second LL ($2 < \nu < 4$) are different than those in the lowest LL
($\nu<2$).
This state is thought to arise from a $p$-wave pairing of CFs described by the Moore-Read Pfaffian \cite{moore91}.
In contrast to the CF hierarchy states, the $\nu=2+1/2$ FQHS is predicted to have quasiparticles that obey 
exotic non-Abelian braiding statistics and which might be harnessed for 
topological quantum computation \cite{dassarma05}. 
However, the nature of this state is not yet settled and it is the subject of an intense investigation 
\cite{willett09,radu08,dolev08,xia04,pan08,pan99,eisen02,miller07,nuebler10,dean08,zhang10}.

An equally interesting and related problem is the origin of the {\it odd denominator} FQHS of the second LL
such as the ones observed at  $2+1/3$, $2+2/3$, $2+4/5$, and $2+2/5$.
While at first blush these states would seem to belong to the CF hierarchy, there is an
increasing body of theoretical work suggesting a more complicated picture.
Studies at $\nu=2+1/3$ find a good overlap of Laughlin's wavefunction 
with the exact numerical solution 
when the finite sample width \cite{peterson08,papic09} or residual interactions between the 
CFs \cite{toke05} are included in the models.
Other theories find that certain odd denominator FQHS of the second LL
might have generalized Pfaffian-like correlations inherited from the nearby $\nu=2+1/2$ FQHS and 
might therefore be fundamentally different from the conventional CF hierarchy states
\cite{read99,bishara08,bonderson08,bernevig08,simon07,levin09,wojs09,simion08}. 
Of these the Read-Rezayi parafermion proposal for the $2+2/5$ state \cite{read99}
is of special importance since, as opposed to the $2+1/2$ Pfaffian, this state ensures
a completeness of the operator space and therefore supports universal topological quantum computation \cite{freedman02}.

It was recently conjectured that nonconventional FQHS form in the $2+1/3 < \nu < 2+2/3$ range of the second LL \cite{wojs09,simion08}.
Within this interesting region so far only a single odd denominator FQHS has been observed at  
$\nu=2+2/5$ \cite{xia04,pan08}. Moreover, convincing signatures of this state such as the observation of a quantized Hall plateau 
and of activated magnetotransport so far come from a single sample \cite{xia04,pan08}. 
Experimental results on the odd denominator FQHS in this range of filling factors can be summarized as follows:
a) the gap of $2+1/3$ FQHS is unexpectedly larger than that of the $2+2/3$ FQHS
\cite{pan99,eisen02,miller07,nuebler10,pan08,dean08,zhang10}, by about a factor two
in the most common samples with densities close to $3.0 \times 10^{11}$~cm$^{-2}$
\cite{pan99,eisen02,miller07,nuebler10,pan08}, and 
b) the ratio of the gaps of the $2+1/3$ and $2+2/5$ FQHS is different than that of
their lowest LL counterparts at $\nu=1/3$ and $2/5$ \cite{xia04,choi08}.
Thus both the even and the odd denominator FQHS of the second LL continue to challenge our understanding.

In this Letter we report the observation of a new FQHS in the second LL at $\nu=2.463 \pm 0.002$ and 
we confirm the existence of the $2+2/5$ FQHS. This new FQHS develops very close to the even denominator $2+1/2$ FQHS 
and from a comparison with the CF hierarchy values we identify it with the odd denominator $\nu=2+6/13$ FQHS. 
Our analysis of the energy gaps of the $2+1/3$ and $2+2/3$ FQHS shows their consistency with the predictions of the MNCF. 
In contrast, the gaps of the $2+2/5$ and that of the newly observed $2+6/13$ FQHS are significantly larger than the values expected using the 
MNCF. This discrepancy of the gaps constitutes the first evidence that
the $2+2/5$ and $2+6/13$ FQHS are not similar to their lowest LL counterparts at $2/5$ and $6/13$ but they are of exotic, possibly non-Abelian nature.

We measured a 4$\times$4 mm$^2$ piece of a 30~nm wide GaAs/AlGaAs 
 quantum well with a density $n=3.0\times 10^{11}$cm$^{-2}$ and mobility $\mu=32\times 10^6$cm$^2$/Vs. 
In order to achieve low charge carrier temperature the sample has been soldered onto
eight sintered silver heat exchangers which were immersed into a liquid He-3 bath \cite{xia04}. 
The temperature of the bath is inferred from the temperature dependent viscosity of the He-3 
which is measured using a quartz tuning fork viscometer \cite{samkhar10}.

\begin{figure}[t]
 \includegraphics[width=1\columnwidth]{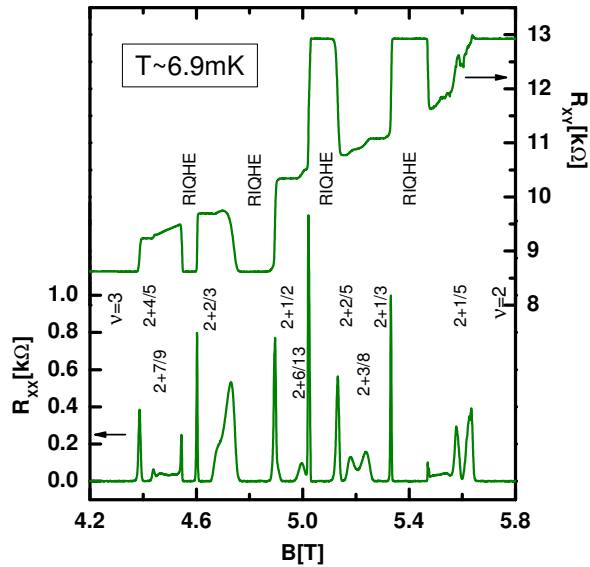}
 \caption{\label{f1}
Magnetotransport for the lower spin branch of the second LL at 6.9~mK. We marked the various FQHS we observe by their LL filling factor 
and the four reentrant integer quantum Hall states by RIQHE.
}
\end{figure}

Figure 1 shows an overview of the longitudinal magnetoresistance $R_{xx}$ and Hall resistance $R_{xy}$
for $2< \nu< 3$. The sharpness of the $R_{xx}$ peaks reveals the quality of the prepared state. 
We observe the prominent FQHS at 
$\nu=2+1/2$, $2+1/3$, and $2+2/3$. In addition we also see an extremely well developed $2+2/5$ state and a less developed but still 
strong $2+3/8$ FQHS. This is the second unambigous identification of these latter states \cite{xia04,pan08} and hence we confirm their existence.
Our ability to cool the sample is evident in the presence of the four fully developed reentrant integer quantum Hall states 
(RIQHS) \cite{eisen02} with wide plateaus \cite{xia04,pan08}. 

\begin{figure}[b]
 \includegraphics[width=1\columnwidth]{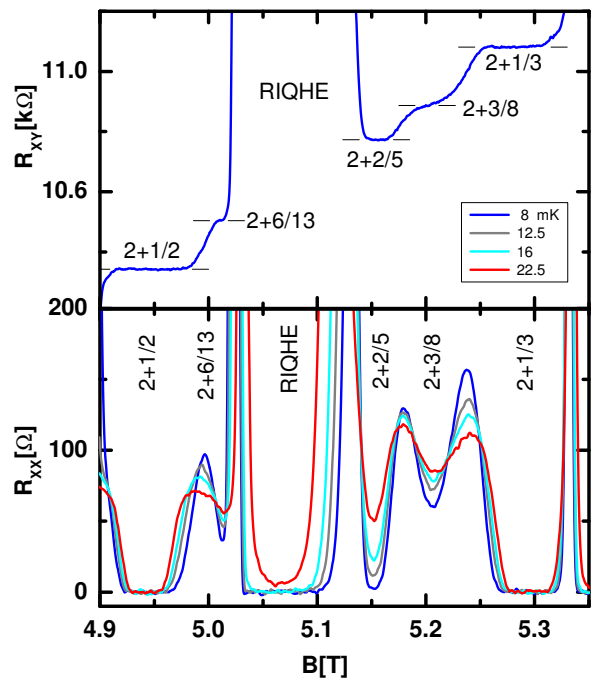}
 \caption{\label{f2}
A magnified view and the temperature dependence of R$_{xx}$ for the fragile FQHS at $\nu=2+6/13$, $2+2/5$, and $2+3/8$. 
The horizontal lines mark the expected quantized values of R$_{xy}$ for each FQHS as referenced to that of the $2+1/2$ FQHS.
}
\end{figure}

We also observe, for the first time, a new FQHS in the very narrow field region between the $\nu=2+1/2$ FQHS and the RIQHS
at slightly higher $B$-fields. This FQHS, shown in more detail  in Fig. 2,
is identified from a well developed narrow minimum in $R_{xx}$ at $\nu=2.463 \pm 0.002$ 
and a plateau in $R_{xy}$ at $h/2.461 e^2$ as determined in reference to the $\nu=5/2$ FQHS.
The new state is independent of the crystallographic direction since its signatures 
are seen when the current is passed along any of the four sides of our sample (not shown). 
We identify this new state with a FQHS at the closest CF hierarchy value $\nu = 2+6/13$.  
The newly seen $2+6/13$ and the previously reported $2+2/5$ FQHS are therefore the 
only odd denominator states observed in the 
interesting region $2+1/3 < \nu < 2+2/3$ where certain theories predict the prevalence 
of generalized Pfaffian-like correlations
\cite{read99,bishara08,bonderson08,bernevig08,simon07,wojs09,simion08}.

Activated transport of the various FQHS is shown in Fig. 3. The energy gaps $\Delta$ extracted from fits of the form 
$R_{xx} \propto \exp(-\Delta/k_B T)$ are also shown. 
We find that in our sample the $2+1/3$ FQHS has the largest gap among the FQHS of the second LL and that
the gaps of the $2+1/3$, $2+1/2$, $2+2/5$, and $2+3/8$ FQHS have reached record values \cite{xia04,pan08,choi08}.
The activated magnetoresistance of the fragile FQHS at $\nu=2+2/5$, $2+3/8$ and $2+6/13$ indicates that the electron temperature follows that of the He-3 bath to the lowest temperatures.

\begin{figure}
 \includegraphics[width=1\columnwidth]{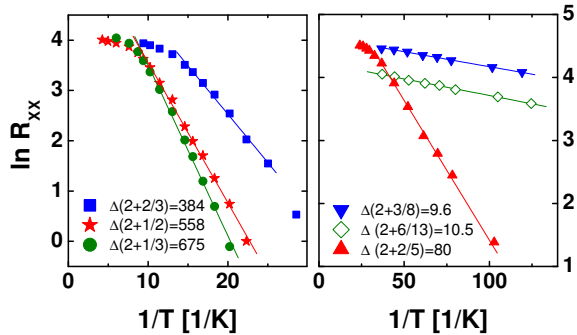}
 \caption{Arrhenius plots for the minimum of R$_{xx}$[$\Omega$] for the prominent 
 FQHS at $\nu=2+1/2,2+1/3,2+2/3$ (left panel) and
 for the more fragile FQHS at $\nu=2+2/5,2+3/8,$ and $2+6/13$ (right panel). 
 Labels indicate the energy gaps extracted in units of mK.
 \label{Fig3}}
 \end{figure} 
 
We now recall quantitative aspects of the MNCF \cite{jainCF}.
Becasue of the flux attachment procedure the CFs move in an effective magnetic field which 
in the vicinity of $\nu=m+1/2$ can be expressed as $B_{\mathrm{eff}}=(1+2m)(B-B(\nu=m+1/2))$ 
\cite{halp93}. 
The energy gap of the FQHS is interpreted as the cyclotron energy of the CF particles less the disorder broadening
term $\Gamma$
\begin{equation}
\Delta=\hbar e B_{\mathrm{eff}} / \mathrm{m_{eff}}-\Gamma ,
\end{equation}
where $\Gamma$ is assumed to be independent of $\nu$ \cite{halp93,du93,mano94,pan00}. 
Such a linear dependence of $\Delta$ on $B_{\mathrm{eff}}$ has been successfully demonstrated 
on the CF hierarchy states of the lowest LL converging towards $\nu=1/2$ \cite{du93,mano94}
and on states in the low $B$-field region of $\nu=1/4$ \cite{pan00}.
It must be kept in mind that because the FQHS appear solely due to electron-electron interactions
the effective mass $\mathrm{m_{eff}}$ of the CFs is not independent of the electron density $n$
but it scales as $\mathrm{m_{eff}} \propto \sqrt{n}$ \cite{halp93,du93,mano94,pan00}. 

In the vicinity of $\nu=5/2$ after substituting $m=2$ we find $B_{\mathrm{eff}}=5(B-B(\nu=5/2))$.
Because of the scarcity of FQHS in the second LL in Fig. 4 we plot the measured gaps as function of the absolute value of $B_{\mathrm{eff}}$.
The nonlinear functional dependence we find for the measured gaps of FQHS of the second LL is in stark 
contrast to the linear dependence for the CF hierarchy states of the lowest LL \cite{du93,mano94,pan00}. 
This difference indicates that at least some of 
the odd denominator FQHS of the second LL cannot be accounted for by the MNCF.

\begin{figure}[b]
 \includegraphics[width=.9\columnwidth]{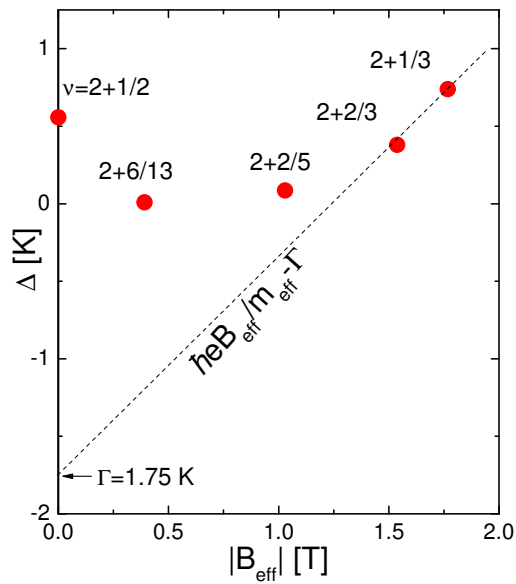}
 \caption{Dependence of the energy gaps $\Delta$ of the FQHS of the second LL with $2+1/3 < \nu < 2+2/3$ on
 the absolute value of $B_{\mathrm{eff}}$. The estimated error for the gaps $\pm 3$\% is less than the size of
 the points. The dashed line is the prediction of the MNCF with $\mathrm{m_{eff}}=0.96 \mathrm{m_{e}}$ 
 described in the text.
 \label{Fig4}}
 \end{figure}
 
To determine which of the FQHS might be a CF hierarchy state we compare our data with the
predictions of the MNCF embodied in Eq. 1 in which we substitute an effective mass 
of the flux two CF of the {\it lowest} LL, i.e. for the FQHS converging to $\nu=1/2$ \cite{du93}.
Because $\mathrm{m_{eff}}$ is slightly different for positive and negative $B_{\mathrm{eff}}$ 
\cite{du93} we average these two different values then apply the
scaling with the density mentioned earlier in this Letter, i.e. 
$\mathrm{m_{eff,1}}/\mathrm{m_{eff,2}}=\sqrt{n_1 /n_2}$ \cite{halp93,du93,mano94,pan00}.
For our sample we obtain $\mathrm{m_{eff}} = 0.96 \mathrm{m_{e}}$, where $\mathrm{m_{eff}}$ is the electron mass in vacuum.
The dashed line of Fig.4 with a slope of $\hbar e / \mathrm{m_{eff}}$ derived from this $\mathrm{m_{eff}}$ 
is in excellent agreement with the gaps of the $2+1/3$ and $2+2/3$ FQHS when $\Gamma=1.75$~K.
Such a value for $\Gamma$ is consistent with values between 1-2~K
obtained from analyses of the gaps of the $\nu=2+1/2$ FQHS \cite{pan99,nuebler10,morf03}
as well as values from the lowest LL \cite{du93,mano94} in high quality samples.
We thus find that the MNCF with a scaled effective mass can account for the gaps of the 
$2+1/3$ and $2+2/3$ FQHS, a result which is interpreted as evidence that these two FQHS 
are part of the CF hierarchy, i.e. are of Laughlin-type.  
We note that this conclusion on the $2+1/3$ and $2+2/3$ states differs from that in Ref.\cite{choi08}.
We find that the so far unexplained large ratio of the gaps of the $2+1/3$ and $2+2/3$ FQHS 
\cite{pan99,eisen02,miller07,nuebler10,pan08} from point a) of the introduction, 
which is also observed in our sample, is a consequence of these two states being Laughlin-correlated.

The plausibility of our argument that the $2+1/3$ FQHS is of Laughlin type is strengthened by the 
following two results. First, our estimated intrinsic gap
$\Delta^{int}(2+1/3)=\Delta+\Gamma \simeq 2.43\;\text{K}=0.021 \; e^2/4 \pi \epsilon l_B$
compares favorably to $0.020 \; e^2/4 \pi \epsilon l_B$, the result of numerics for 
the $2+1/3$ Laughlin state with Coulomb interactions \cite{morf98} and the result for 
the roton gap of a CF model in which the interactions are incorporated through mixing of the LL of the CFs \cite{toke05}. Here $l_B=\sqrt{\hbar /eB}$.  Second, in a recent calculation 
an excellent overlap of the Laughlin wavefunction and the exact numerical solution 
is found at $\nu =2+1/3$ with finite thickness effects included \cite{papic09}.

Figure 4 also shows that Eq. 1 yields negative gaps at $B_{\mathrm{eff}}$ corresponding
to $\nu=2+2/5$ and $2+6/13$ which means that within the MNCF no FQHS are expected to
develop at these filling factors. The FQHS we observe at $2+2/5$ and $2+6/13$ must therefore be of nonconventional origin.
A similar conclusion is reached for the $2+2/5$ in numerical calculations
\cite{read99,bonderson08,scarola00b,wojs09}.
Using the estimated disorder broadening we obtain the intrinsic gaps of these states 
$\Delta^{int}(2+2/5) \simeq 0.0161 \; e^2/4 \pi \epsilon l_B$ and 
$\Delta^{int}(2+6/13) \simeq 0.0157 \; e^2/4 \pi \epsilon l_B$. 

There are several theories which predict series of odd denominator FQHS in the second LL which are different
from the CF hierarchy states \cite{read99,bishara08,bonderson08,bernevig08,simon07,levin09}. 
The Read-Rezayi parafermion theory involving clusters of $k$ electrons
accounts for the $2+2/5$ FQHS through the particle-hole conjugate of the $k=3$ state
but cannot accommodate the $2+6/13$ FQHS \cite{read99,bishara08}. 
The Bonderson-Slingerland hierarchy theory which starts by pairing
charge $e/4$ non-Abelian quasiparticles of the Pfaffian finds FQHS at both $\nu=2+2/5$ and $2+6/13$ with  diagonal 
elements of the coupling-constant matrix $K{_{22}}=-2$ and $K_{22}=-6$, respectively \cite{bonderson08}. 
However, this theory has the disadvantage that the $2+6/13$ FQHS has an unusually high order
$K_{22}=-6$ and states of intermediate $K_{22}$ values are not seen because of the prevalence of the RIQHS. 
The situation is similar for the states constructed using the Jack polynomials \cite{bernevig08}.
The Levin-Halperin theory 
derives the $2+6/13$ state from the anti-Pfaffian in a single step, but it cannot capture 
the $2+2/5$ FQHS \cite{levin09}. 
Taken together, the nature of the $2+2/5$ and $2+6/13$ FQHS remains uncertain. It appears that no single theory
can account for both of them in a natural way and therefore we surmise that these two states have fundamentally different origins.
In the absence of predictions of gaps in the above mentioned theories we cannot further elaborate on the nature of these two states.
Nonetheless, establishing the nonconventional nature of the $2+2/5$ and $2+6/13$
FQHS is a first step towards the understanding of the odd denominator FQHS in the second LL. 

In conclusion, we report the observation of a new FQHS of odd denominator at $\nu=2+6/13$ in the second LL level of a 2D electron gas. 
Our analysis of the energy gaps in terms of the predictions of the MNCF provides 
evidence that in our sample the $2+6/13$ and the $2+2/5$ FQHS do not belong 
to the CF hierarchy and therefore are of exotic nature. The $2+1/3$ and $2+2/3$ FQHS are found, however,
to be consistent with the MNCF which provides a natural explanation for the 
measured ratios of their gaps in numerous samples.
The demonstration of the nonconventional nature of the 2+2/5 state is an important milestone in our understanding which points towards the possible
implementation of universal topological quantum computation with this state.

\begin{acknowledgments}
We thank P. Bonderson, J.K. Jain, and S.H. Simon for insightful comments.
G.A.C. was supported on NSF DMR-0907172 and M.J.M. acknowledges the support of the Miller Family Foundation. 
\end{acknowledgments}


\end{document}